\documentclass[11pt]{article}
\usepackage{amssymb,dsfont,amsmath,amsfonts,hyperref}
\usepackage[active]{srcltx}
\usepackage{slashed}
\usepackage{color}
\advance \textheight by \topskip
\let\oldsqrt\sqrt
\def\sqrt{\mathpalette\DHLhksqrt}
\def\DHLhksqrt#1#2{%
\setbox0=\hbox{$#1\oldsqrt{#2\,}$}\dimen0=\ht0
\advance\dimen0-0.2\ht0
\setbox2=\hbox{\vrule height\ht0 depth -\dimen0}%
{\box0\lower0.4pt\box2}}

\def\be{\begin{equation}}
\def\ee{\end{equation}}
\def\w{\wedge}

\def\e{\epsilon}

\def\s{\star}

\def\mfg{\mathfrak{g}}

\begin{document}

\vspace{7mm}

\begin{center}
{\Large\bf Topological self-dual vacua of deformed gauge theories}
\vspace{1.5cm}

Julio Oliva
\& Mauricio Valenzuela

\vspace*{5mm}

\textit{Instituto de Ciencias F\'isicas y Matem\'aticas, Facultad de Ciencias, Universidad Austral de Chile, Valdivia, Chile}
\end{center}

\vspace{1cm}

\begin{minipage}{.90\textwidth}

\begin{abstract}
We propose a deformation principle of gauge theories in three dimensions that can describe topologically stable self-dual gauge fields, i.e., vacua configurations that in spite of their masses do not deform the background geometry and are  locally undetected by charged particles. We interpret these systems as describing boundary degrees of freedom of  a self-dual Yang-Mills field in $2+2$ dimensions with mixed boundary conditions. Some of these  fields correspond to Abrikosov-like vortices with an exponential damping in the direction penetrating into the bulk. We also propose generalizations of these ideas to higher dimensions and arbitrary p-form gauge connections.
 \end{abstract}

\end{minipage}

\tableofcontents

\section{Introduction}

A self-dual Abelian vector field theory in odd dimensions was proposed in \cite{Townsend:1983xs} and it was later shown  in \cite{Deser:1984kw,Ilha:2001he} that in three dimensions it and its non-abelian generalization are dynamically equivalent to gauge theories with a topological mass  \cite{Siegel:1979fr,Schonfeld:1980kb,Deser:1981wh}. Self-duality in three dimensions can be obtained by means of a Kaluza-Klein compactification from the four-dimensional self-dual condition\footnote{This result is 	 attributed to reference 11 of the paper  \cite{Townsend:1983xs}.}.
 The notion of self-duality \cite{Townsend:1983xs} was extended to gravity in \cite{Aragone1986} and it was shown that it also describe massive spin $2$ degrees of freedom.

 Though self-dual fields in four dimensions are essentially topological objects, in three dimensions, as they where introduced in \cite{Townsend:1983xs}, do not have this property. This is due to their non-vanishing energy-momentum tensor which deforms the background geometry within the context of general relativity. The theories so far discussed may provide models of massive interactions which may dominate at short distances but,  for the above reasons, they are not suitable to describe strongly coupled phases since they cannot be regarded as excitation of any non-perturbative vacuum.  In the present paper we propose gauge systems where, in contrast, the self-dual  fields appear as true vacua configurations.

  By means of a prescription for the deformation of gauge theories in three dimensions, which preserves the gauge symmetries, the self-dual fields  will enjoy a number of properties that are intrinsic to the deformation procedure:

\begin{itemize}
\item[1.-] they will have a self-dual vacua, meaning that they will have non-vanishing gauge curvature but a vanishing action functional,

\item[2.-] their Rosenfeld energy-momentum tensor will vanish,

\item[3.-] their coupling to matter will vanish.

\end{itemize}

Property 1 implies the maximization of the Feynman statistical weight for self-dual electromagnetic fields. Hence the theories here constructed may describe spontaneous magnetization of a medium. The property 2 implies that the self-dual fields will not deform the background geometry, as their will not source the Einstein equations. This is a remarkable since the self-dual fields have non-vanishing masses. The property 3 indicates that the self-dual fields will not affect the dynamics of charged fields, which will behave as free, thought geometrical phases will be produced.

We shall see that the self-dual equation in $2+1$ dimensions can be regarded as a mixed boundary condition for a Yang-Mills field in four dimensions. Indeed, we will show that Abrikosov-like vortices with an exponential damping in the direction penetrating into the bulk will satisfy the self-duality condition in a $(2,2)$ signature bulk space with self-dual boundary conditions. One can imagine that the self-dual boundary conditions allow the end of the vortices to have non-trivial momentum in the boundary, hence they can propagate.

The deformation prescription will in general generate higher derivative extensions. An example of them is the deformation of the Chern-Simons theory, which  turns out to be equivalent to the higher derivative Chern-Simons theory of \cite{Deser:1999pa} with tuned parameters. As a corollary, for these tuned parameters the theory in \cite{Deser:1999pa} will contain self-dual solutions enjoying the  aforementioned properties.

Beside the self-dual fields the deformation procedure may generate other massive degrees of freedom whose dynamics will depend on the details of the model. Hence, the deformation prescription can be regarded as useful for the generation of massive degrees of freedom.

 The paper is organized as follows.  In section 2 we present the deformation method for gauge theories in three dimensions and solve the self-duality constraint for elements in the Cartan subalgebra of the gauge symmetry. In section 3 we apply the deformation to the particular cases of Chern-Simons and BF topological theories. We show that for spin $1/2$ fields the deformation of the minimal coupling term is equivalent to the addition of a Pauli interaction term. In section 4 we will show that the self-duality constraint in $2+1$ dimensions can be  regarded as mixed boundary conditions for self-dual Yang-Mills fields in $2+2$ dimensions. We argue as well that our models may be used to describe the boundary dynamics of an abstract $(2,2)$ signature space-time superconductor. In section 5 we propose extensions of our construction to higher dimensional gauge theories including the $p$-form gauge fields of supergravity in $5$ and $11$ dimensions. Section 6 is dedicated for conclusions and further comments.

\section{Deformation of gauge theories}

Let us consider some generic gauge theory in a three dimensional manifold $M$ whose dynamics is determined by the action principle,
\begin{equation}\label{S0}
S[A]=  \int_{M} \mathcal{L}[A] \, ,
\end{equation}
where the fundamental field $A$ is a gauge connection valued in the Lie algebra\footnote{ We do not refer in particular to supersymmetric theories since our construction is general enough to take them into account, hence the specification to an algebra or a superalgebra is unnecessary.} $\mfg$. Assume that the latter is (quasi)-invariant under the local transformation,
\begin{equation}\label{gt}
A\, \rightarrow \, g^{-1} (A+d)\, g \,, 
\end{equation}
where $g=g(x)$ is an element of the exponentiation of $\mathfrak{g}$.
The variation of the action leads to an expression of the form,
\begin{equation}\label{deltaSgeneric}
\delta S[A]=  \int_{M} \langle \delta A \w \,  \Upsilon[A] \rangle\, + \int_{\partial M} \mathcal{B}[A,\delta A ] \, , 
\end{equation}
where the brackets $\langle\cdot\rangle$ indicate trace, $\Upsilon[ A] $ is a 2-form  constructed from $A$ and its derivatives, and the last contribution stands for a boundary term.  The consequent equation of motion is
\begin{equation}\label{eomundef}
\Upsilon[A] =0 \, ,
\end{equation}
which is naturally solved by the trivial configuration $A=0$. Due to the gauge invariance one must have also that  $\Upsilon[g^{-1} d g]=0$. For an infintesimal gauge transformation, $\delta A = D\epsilon:=d\epsilon+[A,\epsilon] $, the gauge invariance implies from \eqref{deltaSgeneric} that $D\Upsilon[A]=0$.

The addition of a one-form to $A$ transforming as a tensor under the gauge transformations will induce non-linear couplings at the level of the action without spoiling the local symmetry. If there are no other fields involved, the simplest tensor that we can construct is the gauge curvature two-form,  $F:=dA+A^2$, whose Hodge dual provides us with a one-form that can be added to $A$, obtaining thus
\begin{equation}\label{def}
 A^\Theta :=  A + \Theta  \s F \,.
\end{equation}
Here the parameter $\Theta $ has dimensions of length, and the action of the Hodge operator is defined as
$$
\s dx^\mu=\frac{e}{2} \epsilon^\mu{}_{\nu \lambda}  dx^\nu \w dx^\lambda, \qquad \s (dx^\mu \w dx^\nu) = e\, \epsilon^{\mu \nu}{}_\lambda  dx^\lambda, \quad \s \s = \eta\,,
$$
where $\eta$ is the signature of the metric and $e$ is the determinant of frame field $e^a_\mu$.

As expected, the deformed gauge connection \eqref{def} transforms as,
\begin{equation}\label{defgt}
A^\Theta \, \rightarrow \,   g^{-1} ( A^\Theta + d) \, g \,.
\end{equation}

It is convenient  to define the \textit{deformation map},
\begin{equation}
\texttt{def}_\Theta \, : \, \mathcal{O}[A] \quad \rightarrow \quad \mathcal{O}_\Theta[A]:=\mathcal{O}[A^\Theta]\,,
\end{equation}
of functionals of $A$ into functionals of $ A^\Theta $.

In particular,
  \begin{equation}
A^\Theta= \texttt{def}_\Theta \, A\,. 
\end{equation}
Thus, the covariant derivative of a $p$-form $\Omega_p$,
\begin{equation}\label{D}
D \, \Omega_p := d\Omega_p + [A \,, \Omega_p ]\,,
\end{equation}
where $[A  \,, \Omega_p ] := A  \w \Omega_p - (-1)^p \, \Omega_p \w A $,
gets deformed into,
\begin{equation}\label{defD}
D_\Theta \, \Omega_p := \texttt{def}_\Theta D\Omega_p = d\Omega_p + [A^\Theta \,, \Omega_p] \,.
\end{equation}

The deformation of the action functional \eqref{S0} is given then by,
\begin{equation}\label{STheta}
 S_{\Theta}[A]:=\texttt{def}_\Theta \,S[A]= S[A + \Theta  \s F ]\,. 
\end{equation}

$S_\Theta [A]$ contains the undeformed action at zeroth order in $\Theta$, plus higher derivative terms.

Note that pure gauge configurations, $F=0$, are fixed points of the deformation map, while its kernel,
\begin{equation}\label{ker}
ker\{\texttt{def}_\Theta \, A \} := \{ A= A_{sd}  \,\,  | \, \,  \texttt{def}_\Theta \,  A_{sd} = 0 \,\} \, ,
\end{equation}
is by its definition formed by non-abelian self-dual connections, since $\texttt{def}_\Theta \,  A_{sd}=  A_{sd}+\Theta \s F_{sd}$ and hence
\begin{eqnarray}\label{SD}
  A_{sd}+\Theta \s F_{sd}&=&0 \,,
\end{eqnarray}
which is the self-dual condition in three dimensions \cite{Townsend:1983xs}. According to this equation the vanishing of the divergence follows:
\begin{equation}\label{gf}
D \s A_{sd}=2 \partial_\mu (e\,  A^\mu_{sd}) =2 \, e\,  \nabla_\mu A^\mu_{sd} = 0 \, ,
\end{equation}
where $\nabla$ is the Levi-Civita connection.  We can show that these self-dual configurations are massive by acting with the exterior derivative on the expression \eqref{SD}, and then using  \eqref{SD} again. Thus we obtain
\begin{equation}\label{consistencySD}
(  \s d  \s  d - \Theta^{-2} ) A_{sd}  + (\s d \s -\Theta^{-1} \s) \,  A_{sd}  \w A_{sd}=0 \, .
\end{equation}
The first term  is equivalent in a local frame to
\begin{equation}\label{KGsd}
(  \s d  \s  d - \Theta^{-2} ) A_{sd} = dx^\mu ( (\square - \Theta^{-2} ) \delta_\mu{}^\nu - R_\mu{}^\nu) A_{sd\, \nu} - dx^\mu \nabla_\mu \nabla^\nu  A_{sd\, \nu} \,,
\end{equation}
whose last term vanishes on account of the condition \eqref{gf}. The mass of the self-dual field is determined by the first term in \eqref{consistencySD} and is given by $|\Theta^{-1}|$.

We obtain the deformation of the variation of $A$ by Requiring that the deformation and the variation of the it should be  equivalent to the variation of the deformation of $A$, that is
\begin{equation}
\texttt{def}_\Theta  \, \delta A = \delta  A^\Theta \, , \qquad
\delta  A^\Theta= \delta A + \Theta  \s D \, \delta A    \,.
\end{equation}
Hence the operations of variation and deformation of an action principle should commute. Indeed we obtain
\begin{eqnarray}\label{deltadefS0}
 \delta S_\Theta[A] &=&  \int_{M}  \langle (\delta A + \Theta  \s D \, \delta A )\, \w  \Upsilon[A^\Theta] \rangle + \int_{\partial M} \mathcal{B}[A^\Theta ,\texttt{def}_\Theta  \, \delta A ]\hspace{2mm}\\
  &=&\int_{M}  \langle \delta A \w ( 1  + \Theta   D \s ) \Upsilon[A^\Theta] \, \rangle + \int_{\partial M} (\mathcal{B}[A^\Theta ,\texttt{def}_\Theta  \, \delta A ] + \Theta \langle \delta A \w \s \Upsilon[A^\Theta] \rangle)\,,\nonumber
 \end{eqnarray}
 which provides the deformation of the variation of the action functional $S[A]$ given in \eqref{deltaSgeneric}. Hence the equations of motion of the deformed theory read
\begin{equation} \label{defeom1}
 ( 1  + \Theta   D \s ) \Upsilon_\Theta[A]  =0 \, , \qquad \Upsilon_\Theta[A]:=\Upsilon[A^\Theta] \, .
\end{equation}
Since $A=0$ is a  trivial solution of \eqref{eomundef}, i.e. $\Upsilon[0]=0$, the self-dual field $A_{sd}$, for which \eqref{SD} holds, is solution of \eqref{defeom1} since $\Upsilon_\Theta[A_{sd}]=\Upsilon[A_{sd}+\Theta \s F_{sd}]=\Upsilon[0]=0$.  The value of the action for these configurations vanishes, as it is equivalent to evaluating the undeformed action functional in the trivial value of its argument $A=0$,
\begin{equation}
S_\Theta [A_{sd}]=S [ A_{sd}+\Theta \s F_{sd}]=S[0]=S_\Theta [0]=0\,.
\end{equation}
The Rosenfeld energy-momentum tensor obtained from the variation of the action with respect to the background metric, ignoring boundary terms, is given by
\begin{eqnarray}\label{deftmunu}
\frac{\delta\;}{\delta g^{\mu \nu}} S_\Theta[A] &=& T_{\mu \nu}[A+\Theta \s F]+ \frac{\delta A^\Theta\;}{\delta g^{\mu \nu}}  \frac{\delta\;}{\delta A^\Theta} S_\Theta[A] \\
&=& T_{\mu \nu}[A+\Theta \s F]- \frac{\Theta}{2} g_{\mu \nu} \langle \Upsilon_\Theta[A] \s F \rangle \, ,
\end{eqnarray}
where
$$
T_{\mu \nu}[A+\Theta \s F] = \mathtt{def}_\Theta \, T_{\mu \nu} [A]\, ,
$$
is the energy momentum tensor of the undeformed theory \eqref{S0},
$$
T_{\mu \nu}:=\frac{\delta\;}{\delta g^{\mu \nu}} S[A] \,,
$$
valued in the deformed connection.

Since, for self-dual solutions, equation \eqref{SD} and $\Upsilon_\Theta[A_{sd}]=0$ hold, the deformed energy-momentum tensor \eqref{deftmunu} vanishes,
\be
\frac{\delta\;}{\delta g^{\mu \nu}} S_\Theta[A] \Big|_{A=A_{sd}}=T_{\mu \nu}[0] =0 \, .
\ee
Hence, the self-dual solution of the deformed theory do not source gravity, in the sense that the field equations for the spacetime metric $g_{\mu\nu}$ will be the same as in the trivial vacuum $A=0$.

The self-duality equation \eqref{SD} is not gauge invariant by itself, but the deformed equation of motion \eqref{defeom1} are indeed gauge invariant since for them
\be\label{sdgauged}
\Upsilon_\Theta [A^g_{sd}] = \Upsilon [A^g_{sd}  + \Theta \s F^g_{sd}  ] = \Upsilon [g^{-1}(A_{sd}  + \Theta \s F_{sd} + d) g  ]= \Upsilon [g^{-1} \, d \, g  ]=0\,.
\ee
 Thus, gauge transformations acting on self-dual solutions may generate non self-dual configurations which are still in the spectrum of the theory.

It is straightforward to see that the Feynman statistical weight constructed with the deformed action takes the same values  for the trivial and for self-dual configurations,
\begin{equation}
\exp (i S_\Theta[A]) |_{A=0}= \exp (i S_\Theta[A])|_{A=A_{sd}}=1\,.
\end{equation}
Thus the vacua of a deformed system can be either voided or filled by  self-dual electromagnetic fields, where they could appear as forming magnetic domains. This feature may be of use for the description of two dimensional layers of materials exhibiting spontaneous magnetization.

\subsection{Vortex solutions}\label{secvortex}

In order to find some solutions of the self-dual condition let us look for $A$ in Minkowski space-time and $A \, \in \mathfrak{g}_0\otimes T^*(M)$, where  $\mathfrak{g}_0 \subset \mathfrak{g}$ is an Abelian subalgebra of the gauge algebra. The self-dual equation \eqref{SD} now becomes,
\be \label{aveq1}
A + \Theta \s dA =0 \,, \qquad A \in \mathfrak{g}_0 \, ,
\ee
which together with the static ansatz, $\partial_0 A_{ \mu}=0 $, yields,
\be\label{aveq2}
A_{0}+\Theta \e_{ij} \partial_i A_{ j}=0\, ,\qquad A_{ i}-\Theta \e_{ij} \partial_i A_{v\,0}=0 \, ,
\ee
where we define the antisymmetric tensor as $\e_{ij}:=\e_{0ij}$, $i,j=1,2$. The solutions of these equations are given by
\be \label{avsol1}
A=(-\Theta B, \Theta^2 \partial_2 B , - \Theta^2 \partial_1 B ) \, ,\qquad \mu=0,1,2,
\ee
such that $B$ satisfies $B-\Theta^2 \partial_i \partial^i B =0 ,$ i.e.
\be \label{avsol2}
B=B(x^1,x^2)=B(r \, sin\, \vartheta, r \, cos \, \vartheta)= \sum_{ \ell}  \,  c_{\ell}  \, T \, \exp ( \textit{i}\,  \ell  \, \vartheta ) \; K_{\ell} (\tfrac{r}{|\Theta|} ) \, ,  \qquad \, T \, \in \,  \mathfrak{g}_0 \, .
\ee
Here $(r,\vartheta)$ are the polar coordinates of the spatial plane, $\ell \in \mathbb{Z}$, $\, c_{\ell}$ are integration constants and $K_{\ell } (r/ \Theta )$ is a Bessel function of the second kind, hence vanishing at spatial infinity. Note that the magnetic field decays as an Abrikosov vortex does in the radial direction, with London penetration depth $\Theta$.
The holonomy of the connection defined by the vortex field  \eqref{avsol1} is,
\be \label{holonomy}
\exp \oint A = \exp \int_\Sigma F = \exp  \Phi
\ee
where $\Sigma$  is the surface inside the contour of integration and $\Phi$ is the  magnetic flux. Here  we have used Stokes' theorem and that $A\w A$ vanishes. When the contour is taken to infinite we obtain that,
\be \label{phi0}
\Phi_0=\int_{\mathbb{R}^2} B\, r dr d\vartheta T= 2\pi \Theta^2  c \, T \, ,
\ee
from where we observe that only the $\ell=0$ mode of \eqref{avsol2} contributes to the integral \eqref{phi0} and we have defined $c:=c_{0}$. Using the latter  \eqref{avsol2} can be written instead as
\be \label{B}
B=\frac{\Phi_0}{2\pi \Theta^2}  \;  \sum_{ \ell}  \,  \tilde{c}_{\ell}\exp ( \textit{i}\,  \ell  \, \vartheta )  \; K_{\ell} (\tfrac{r}{|\Theta|} ) \, ,  \qquad \, {\Phi_0} \, \in \,  \mathfrak{g}_0 \, ,
\ee
where $\tilde{c}_{\ell}:=c_{\ell}/c_0$.
The connection $A$ will be local  if \eqref{holonomy} is the identity, thus imposing  the restriction
\be
\Theta^2  c \, T=i N \,, \qquad \, N \in \mathbb{Z},
\ee
on the constant of integration $c$ and the eigenvalues of the generator $T\in \mathfrak{g}_0$. Hence the magnetic flux is  quantized.

Note that vortices \eqref{avsol1} can be linearly superposed, since they are valued in the abelian subalgebra $\mathfrak{g}_0$.

\subsection{Inclusion of matter fields }

We now study the effect of the deformation map in theories with matter fields. For that purpose let us consider the action principle
\begin{equation}\label{gplusm}
S^{tot}[A,\Psi]=S[A]+S^{int}[A,\Psi] + S^{m}[\Psi] \, ,
\end{equation}
where
\begin{equation}\label{mincoup}
S^{int}[A,\Psi]  = - \int_{M}   \langle A \w \s J[\Psi] \rangle \, ,
\end{equation}
is the minimal coupling term and $J[\Psi]$ is the current one-form associated to the field $\Psi$, whose free action is $S^{m}[\Psi]$.

The variation of \eqref{gplusm} with respect to $A$ yields
\begin{equation} \label{undefeommc}
 \Upsilon[A]  =\s J[\psi] \, ,
\end{equation}
where we have made use of \eqref{deltaSgeneric}.

Let us now deform \eqref{gplusm} as prescribed, i.e. let us obtain $ S_\Theta[A,\Psi]:=\texttt{def}_\Theta \, S[A,\Psi]$,
\begin{equation}\label{defgplusm}
S^{tot}_\Theta [A,\Psi]=S_\Theta [A] + S^{int}_\Theta [A ,\Psi] + S^{m}[\Psi]\,,
\end{equation}
where,
\begin{equation}\label{gaugemattercoup}
 S^{int}_\Theta [A ,\Psi]=S^{int} [A+\Theta \s F ,\Psi] =-\int_{M}   \langle A \w \s J + \eta \Theta  \ F \w  J\, \rangle \,.
\end{equation}
and we have used $\s\s= \eta$.
 Thus the deformation generates the coupling of the matter current with the field strength. Indeed, when the matter consists of Dirac fermions the deformed coupling corresponds to the Pauli term in electrodynamics, with $\Theta$ playing the role of the dipole moment. This will be shown in section \ref{defdirac}.

The variation of the interaction term with respect to $A$ yields
 \begin{equation}\label{mincouppsi}
\delta S^{int}_\Theta [A ,\Psi]=-\int_{M}   \langle \delta A \w (\s J + \eta \Theta  D J ) \rangle - \int_{\partial M} \langle \eta \Theta \delta A \w J \rangle \,,
\end{equation}
  which together with the variation of the gauge field term \eqref{deltadefS0} yields the equation of motion
\begin{equation} \label{eomdefgaugemat}
 ( 1  + \Theta   D \s ) (\Upsilon[A^\Theta]  -\s J)=0  \, .
\end{equation}
Taking the covariant derivative of this expression one obtains
\begin{equation} \label{defconsc}
 D_\Theta \s (   J -\eta \s \Upsilon[A^\Theta]) = 0 \,,
\end{equation}
where $D_\Theta$ is the deformed covariant derivative \eqref{defD}.
Evaluated on a self-dual $A$ the equation of motion \eqref{eomdefgaugemat} gets reduced to
\begin{equation}\label{Jeq}
(\s J +\eta \Theta D J)|_{A=A_{sd}} =0 \,,
\end{equation}
since $\Upsilon[A^\Theta=0]=0$, while \eqref{defconsc} yields the conservation of the current,
\begin{equation}\label{Jcons}
d \s J |_{A=A_{sd}} = 0 \,.
\end{equation}

As an additional remark, we can also consider the coupling of $A$ with a two-form field, say $C$, with interaction term   $S^{int}[A,C]  = - \int_{M}   \langle A \w C \rangle $. The consequent equations of motion coming from the variation of $A$ can be obtained by formally replacing $\s J=C$ in \eqref{eomdefgaugemat}-\eqref{Jcons}, while $C$ must be given an action term for itself.

\section{Applications of the deformation map}

Although it is not gauge invariant, it is illustrative to see the effect of the deformation acting on a Proca mass term for a vector field,
\begin{equation}\label{void}
 - m \, \int_{M} \langle \, A\w \s A \, \rangle \,.
\end{equation}
 Here $m$ is a constant with dimensions of mass and we assume a Lorentzian signature. Its deformation is
\begin{equation}\label{defSvacuum1}
- m \, \int_{M} \langle \, A^\Theta\w \s A^\Theta \, \rangle \, =  m \int_{M} \langle - \, A\w \s A + 2 \Theta  A\w  F + \Theta^2 F\w \s F \rangle \, ,
\end{equation}
which is equivalent to the extended self-dual theory introduced in \cite{Deser:1984kw} with fixed relative couplings. We can take the limit
\be \label{limit}
m \Theta^2=\frac{1}{4 \mathtt{e}}\,, \qquad \Theta  \quad \rightarrow \quad \infty\, , \qquad m  \quad \rightarrow 0\, ,
\ee
where $\mathtt{e}$  is a constant, and obtain from \eqref{defSvacuum1}   the Yang-Mills action,
\be \label{YMlimit}
\frac{1}{4 \mathtt{e}}\,\int_{M} \langle F\w \s F \rangle \quad \leftarrow \quad - m \, \int_{M} \langle \, A^\Theta\w \s A^\Theta \, \rangle \,  \,.
\ee
We observe here that the coupling constant of the resulting Yang-Mills action is related to the vanishing mass limit of the self-dual field described by  \eqref{defSvacuum1}, $\Theta^{-1}$, and the mass constant coming from a Proca action in the strongly coupled limit described by \eqref{void}.

 The above example suggest that the deformation prescription provides us with interpolations between two regimes of a non-Abelian vector system, either very heavy and non-propagating or propagating and massless. We will observe similar features in the following deformation of two topological theories.

\subsection{Deformation of Chern-Simons theory}

Let us write down the basic equations of  Chern-Simons theory.
The action is given by
\begin{equation}\label{Scs}
S_{CS}[A]= \frac{\kappa}{4\pi} \int \left\langle AdA+\frac{2}{3} A\w A \w A \right\rangle \,,
\end{equation}
and its  variation yields
\begin{equation}\label{deltaS}
\delta S_{CS}[A]=\frac{\kappa}{2\pi} \int_M \langle \delta A \w \,F \rangle -\frac{\kappa}{4\pi} \int_{\partial M } \langle A \delta A \rangle \,,
\end{equation}
with the subsequent  equation of motion,
\begin{equation} \label{F}
F=dA+A\w A=0\, .
\end{equation}

The Chern-Simons action \eqref{Scs} is deformed into
\begin{equation}\label{defS}
\texttt{def}_\Theta \, S_{CS} [A]= S_{CS\, \Theta}[A]=S_{CS}[A]+ S_{YM}[A] + S_{ext}[A] + S_{\partial M} \, ,
\end{equation}
consisting of a Yang-Mills action
\begin{equation}\label{YM}
S_{YM}[A]= \frac{\Theta   \kappa}{2\pi} \int \langle F \w \s F \rangle\,,
\end{equation}
and the higher derivative term
\begin{equation}\label{extS}
S_{ext}[A]= \frac{ \Theta ^2 \kappa}{4\pi} \int  \left\langle  \s F \w D(\s F) + \frac{2\Theta }{3} \s F \w \s F \w \s F \right\rangle  \,,
\end{equation}
plus the boundary term,
\begin{equation}\label{YM}
S_{\partial M}[A]=  - \frac{\Theta  \kappa}{4\pi} \int \langle d (A\w \s F)  \rangle \,.
\end{equation}
We can see that for small values of $\Theta $, at linear level, and ignoring the boundary term,  one obtains the  topologically massive gauge theory studied in \cite{Siegel:1979fr,Schonfeld:1980kb,Deser:1981wh}, with mass $\Theta ^{-1}$.

The equations of motion are obtained from the variation of the deformed action,
\begin{eqnarray}\label{defEOM1}
\delta \, \texttt{def}_\Theta  S_{CS}[A] &=& \frac{\kappa}{2\pi} \int_M \langle \delta A^\Theta \,F_\Theta \rangle -\frac{\kappa}{4\pi} \int_{\partial M } \langle A^\Theta \delta A^\Theta \rangle\,, \\[5pt]
&=& \frac{\kappa}{2\pi} \int_M  \langle\delta A  \,(\Theta  D (\s F_\Theta ) + F_\Theta ) \rangle \nonumber \\[5pt]
&& -\frac{\kappa}{4\pi} \int_{\partial M } \langle  A^\Theta (\delta A + \Theta  \s D\, \delta A) \rangle\, ,
\end{eqnarray}
where
\begin{equation}\label{defF}
F_{\Theta}:=\texttt{def}_\Theta F= F+\Theta  D \s F+\Theta ^2 \s F \s F \,.
\end{equation}
Thus the equations of motion read
 \begin{equation}\label{EOM}
( 1+\Theta  D  \s ) F_\Theta=( 1+\Theta  D  \s )(d+A^\Theta) A^\Theta =0  \, ,
\end{equation}
which is in agreement with \eqref{defeom1}.

The action \eqref{defS} is equivalent to the higher derivative extension of Chern-Simons theory proposed in \cite{Deser:1999pa} with tuned parameters. One can verify that for a self-dual field the energy-momentum tensor of this theory vanishes. We should also comment that at first order around $\Theta\approx 0$, considering the coupling to matter described in the previous section, one obtains the system proposed in reference \cite{Itzhaki:2002rc} for the description of anyons.

\subsection{Deformation of  BF theory}

The BF theory,
\begin{equation}\label{S}
S_{BF}[A,\mathcal{B}]=  \int \left\langle \mathcal{B} \w F \right\rangle \,,
\end{equation}
where $\mathcal{B}$ is a 1-form and $F$ is the curvature of the gauge connection $A$, is a topological theory whose equations of motion read,
\be \label{eomBF}
F=0\, , \qquad D \mathcal{B}=0 \,.
\ee
Let us deform $A$ as usual, obtaining thus the deformed BF action,
\begin{equation}\label{S}
S_{\mathcal{B}F\, \Theta }[A,\mathcal{B}]= \int \left\langle \mathcal{B} \w F_\Theta \right\rangle \,,
\end{equation}
where  $F_\Theta$ is defined in \eqref{defF}. Now the equations of motion read,
\be \label{defeomBF}
F^\Theta = 0 \,, \qquad (1+ \Theta D \s ) D_\Theta \mathcal{B} =0 \, .
\ee
A first simple solution of these equations is given by pure gauge configurations,
$$
A=g^{-1} d g\,, \quad \mathcal{B}=g^{-1} \mathcal{B}_0 g \,,
$$
where $g$ is an element of the gauge group and $\mathcal{B}_0$ is a constant, i.e. $d\mathcal{B}_0=0$. One can also construct a second solution for a self-dual $A$ for which $F^\Theta=0$. Then imposing $[\mathcal{B},A]=0$ we are left with a massive spin $1$ vector field equation,
$$
(1+ \Theta d \s )d\mathcal{B}=0\,,
$$
which is the equation for a topologically massive Abelian field with mass $\Theta^{-1}$ \cite{Deser:1981wh}.

\subsection{Deformation of a gauge theory coupled to spin $1/2$ fermions}\label{defdirac}

Let us now deform the  action of a gauge field  minimally coupled to a spin $\tfrac{1}{2}$ fermion in three dimensions. This means, in the action functional the gauge connection should be replaced by the deformed connection $A_\Theta$, thus yielding
\begin{equation}\label{deffermionaction}
S^{Dirac}[\psi, A^\Theta]=\int d^3x e \, \bar{\psi} \Big(\overleftarrow{ \slashed{D}}_\Theta-\overrightarrow{\slashed{D}}_\Theta \Big) \psi \, , \qquad \overrightarrow{\slashed{D}}_\Theta: = \gamma^\mu (\overrightarrow{\partial}_\mu + A_{\Theta\, \mu}) \,, \quad \overleftarrow{\slashed{D}}_\Theta =  (\overleftarrow{\partial}_\mu - A_{\Theta\, \mu}) \gamma^\mu\,.
\end{equation}
The current one-form is
\be\label{fermion}
J_n=dx^\mu \, \bar{\psi} \, \gamma^\mu \, t_n\,  \psi \,, \qquad t_n \, \in \, \mathfrak{g} \,,
\ee
where $\gamma_\mu$ is a Dirac matrix in the coordinate basis and $t_n$ is a generator of the Lie algebra where the (deformed) gauge connection is valued.
The variation with respect to $\psi$ yields the equation of motion
\be \label{defDiraceq}
\overrightarrow{\slashed{D}}_\Theta \, \psi =0 \, ,
\ee
which in components reads
\be \label{defDiraceq1}
 \Big( \gamma^\mu \overrightarrow{\partial}_\mu + \gamma^\mu A_\mu - \frac{\Theta}{4} [\gamma^\mu, \gamma^\nu] F_{\mu\nu} \Big) \, \psi= 0 \,.
 \ee
 Here we have made use of the equality  $[\gamma_\mu, \gamma_\nu ]= - 2 e \epsilon_{\mu \nu\lambda} \gamma^\lambda$, obtained from the Clifford algebra $\gamma_\mu  \gamma_\nu= g_{\mu\nu} - e \epsilon_{\mu \nu\lambda} \gamma^\lambda$, and hence,
 \be \label{ppl}
 \gamma^\mu A^\Theta_\mu = \gamma^\mu A_\mu - \frac{\Theta}{4} [\gamma^\mu, \gamma^\nu] F_{\mu\nu} \,.
 \ee
  Using that $\nabla^\mu \gamma_\mu=0$, where $\gamma_\mu=e^a_\mu \gamma_a$, and since the $\nabla^\mu$ is the Levi-Civita connection so that $\nabla^\mu \,  e_\mu^a =0$, one can show that the current is conserved, $ D_\Theta \s  J=0$, which together with \eqref{defconsc} yields $D_\Theta \Upsilon[A^\Theta]= 0 $.

We observe in \eqref{defDiraceq1} that the Pauli term appears, with $\Theta$ playing the role of magnetic dipole moment. By definition, for a self-dual connection the expression \eqref{ppl} identically vanishes. Therefore in the presence of such gauge fields the Dirac equation in \eqref{defDiraceq1} describes a locally free fermion, though  geometrical phases may be generated  as it occurs in the Aharanov-Bohm effect.

\section{Topological vortex in a $M^+_{2,2}$ superconductor}

The fact that the three dimensional self-dual fields in the deformed theories are locally undetectable,  by both the geometry and by charged particles, implies that they are stable against deformations of the geometry and the presence of electric charges, suggesting that they are of a topological nature. A way to make sense of them is to think that they are topological objects in a higher-dimensional space-time but with non-trivial boundary degrees of freedom. In order to show that this is indeed the case, we will appeal to the simplest topological system in four dimensions, i.e. self-dual solutions of Yang-Mills theory, and look for configurations that at the boundary satisfy the three dimensional self-duality equation. This is, we propose the system of equations,
\be \label{sd4d}
F_4- \, \s_4 F_4=0 \,, \qquad  \Big( A_4+\Theta \s_3 F_4\Big) \Big\vert_{\partial \mathcal{M}} =0 \,,
\ee
where; $F_4:=d_4 A_4 + A_4 \w A_4$  is the curvature of the gauge field $A_4$ defined in a four dimensional manifold $ \mathcal{M}$, $d_4$, $\s_4$ are respectively the exterior derivative and the Hodge operator defined on $ \mathcal{M}$,  $\vert_{\partial \mathcal{M}} $ stands for the restriction to the boundary, and $ \s_3$ is the Hodge operator of the boundary manifold constructed with the induced boundary metric.

The boundary condition in \eqref{sd4d}, which we may refer  to as Robin's boundary condition since it combines Neumann's and Dirichlet's, is equivalent to the self-duality equation in three dimensions \eqref{SD} with the identifications $A_4\vert_{\partial \mathcal{M}}=A_{sd}$, $\partial \mathcal{M}=M$ and $\s_3=\s$, according to our previous conventions. Since we already know solutions of the self-dual system in three dimensions we may look for solutions of the full system \eqref{sd4d} by separation of variables, i.e. multiplying $A_{sd}$ by a function of the coordinate parametrizing the interior of the bulk space. Doing this and then fixing the boundary geometry to be Lorentzian implies that  non-trivial solutions of \eqref{sd4d} do exist only when the signature of the bulk geometry is $(2,2)$.

Let the bulk space be $\mathcal{M}=M^+_{2,2}:=M_{1,2} \times \mathbb{R}^+$, with signature $(-,+,+,-)$, where $M_{1,2}$ is a Minkowski space-time slice with signature $(-,+,+)$, while  $\mathbb{R}^+ \, \ni \, z$ is a time-like non-negative half-line. The bulk coordinates are labeled by $(x^\mu,z)$, and the boundary $M=\partial M^+_{2,2}=M_{1,2}$ is at $z=0$. Under these considerations we obtain the following solution of \eqref{sd4d}:
\be \label{qflat}
A_4= \exp \Big(- \frac{z}{\Theta}  \Big) A_{sd} \,, 
\ee
where $A_{sd}$ is a solution of \eqref{SD}.

In particular for $A_{sd}=A$ given by \eqref{avsol1} and  \eqref{B}, equation \eqref{qflat} yields the following component of the field strength in the direction of the coordinate $z$
\be \label{B1}
{\bf B}_z:=(F_4)_{21} = \exp \Big(- \frac{z}{\Theta}  \Big)  \frac{\Phi_0}{2\pi \Theta^2}  \;  \sum_{ \ell}  \,  \tilde{c}_{\ell}\exp ( \textit{i}\,  \ell  \, \vartheta )  \; K_{\ell} (\tfrac{r}{|\Theta|} ) \, ,  \qquad \, {\Phi_0} \, \in \,  \mathfrak{g}_0  \, ,
\ee
where $r$, $\vartheta$ and $z$ are the radial, angular and axial cylindrical coordinates respectively.

Expanding the Bessel functions for $r>>\Theta$ one obtains
\be \label{B2}
{\bf B}_z \approx \frac{\Phi_0}{2\sqrt{ 2\pi \Theta r} } \,  \exp \Big(- \frac{z+r}{\Theta}  \Big) \, ,
\ee
which is very similar to the form of an Abrikosov vortex but with an exponential damping of the electromagnetic field into the bulk.

With this explicit solution we have shown that self-dual fields in three dimensional Minkowski space-time can be regarded as propagating  boundary degrees of freedom of a topological self-dual Yang-Mills field in $2+2$ dimensions.

The equation \eqref{qflat} is actually a solution of the more general system,
\be \label{YMR}
D_4 \s_4 F_4=0 \,, \qquad  \Big( A_4+\Theta \s_3 F_4\Big) \Big\vert_{\partial \mathcal{M}} =0 \,,
\ee
which is a Yang-Mills system with Robin's self-dual boundary conditions, where $D_4$ is the gauge covariant derivative defined in four dimensions.

We would like to endow the boundary degrees of freedom with their own dynamics. In order to ensure the topological stability of the system \eqref{sd4d} the boundary theory should not back react on the bulk field equations. Hence, the deformed theories here proposed are suitable for this purpose, and one may consider the following system:
\begin{equation} \label{defeom4d}
 D_4 \s_4 F_4=0 \,, \qquad  ( 1  + \Theta   D_4 \, \s_4 ) \Upsilon_\Theta[A_4] \Big\vert_{\partial \mathcal{M}}  =0 \, ,
\end{equation}
with gauge invariant boundary conditions given in \eqref{defeom1},  which has \eqref{qflat} as a possible solution.

Since self-dual fields provide bounds of the Yang-Mills action in four dimensions, vortices like \eqref{qflat} could appear as vacua configurations of Yang-Mills theory forming condensed states. The cusp shape of the vortex \eqref{B1}-\eqref{B2} suggests that, althought here they are treated as abstract constructions in $2+2$ dimensions, they could in practice describe real vortices in superconductors.

\section{Higher dimensional extensions}

A natural question that may appear after the results so far obtained, is whether analogous deformations of gauge systems could be performed in higher dimensions exhibiting similar features. Here we hint at some possibilities to explore.

The basic idea is to look for suitable functions of the gauge curvature that will transform as tensors under gauge transformations and redefine the gauge connection by its addition. This means, the deformation of a $p$-form $A_{[p]}$ gauge connection will be given by an expression with the following structure:
\be\label{sdgeneral}
A_{[p]}^{\bf \Theta} := A_{[p]}+ f_{[p]}({\bf \Theta}; F_{[p+1]}\, ; \s ) \;.
\ee
The shift $f_{[p]}({\bf \Theta}; F_{[p+1]}\, ; \s ) $ is a function of the set of constant parameters ${\bf \Theta}$, the gauge curvature tensor $F_{[p+1]}$, and the spacetime metric implicit in the Hodge dual operator $\s$.  The form of the shift is restricted by diffeomorphism and gauge covariance. The kernel of the deformation map \eqref{sdgeneral} is defined by the nontrivial solutions of the equation
\be\label{sdgeneral1}
A_{[p]}+ f_{[p]}({\bf \Theta}; F_{[p+1]}\, ; \s )=0 \,.
\ee
For those configurations the deformed action principle $S[A_{[p]}^{\bf \Theta} ]$, obtained from the undeformed action $S[A_{[p]}] $, will vanish and it is expected that the derived Rosenfeld energy momentum tensor and the coupling to other matter fields will vanish.

For instance in  $2n+1$ dimensions one can consider the following deformation,
 \begin{equation}
A^\Theta_{[1]}:= A_{[1]}+\Theta \s (F_{[2]}\w)^n \,,
\end{equation}
which could be used to deform a Yang-Mills and/or a Chern-Simons action.

For an example involving a higher rank gauge connection we can take
\begin{equation}
A^\Theta_{[n]}:=A_{[n]}+\Theta \s F_{[n+1]} \, .
\end{equation}
In this case the kernel of the deformation map  is equivalent to the space of solutions of the higher dimensional  self-dual equations introduced in \cite{Townsend:1983xs},
\be\label{sdodd}
A_{[n]}+\Theta \s F_{[n+1]}=0\,.
\ee

More precise cases of study would be five or eleven dimensional $N=1$ supergravities. In such theories the one-form $A_{[1]}$ and three-form $H_{[3]}$ respectively can be deformed as
\begin{equation}
A_{[1]}\rightarrow A_{[1]}+\Theta \s \left( F_{[2]} \wedge F_{[2]} \right) \;,  \qquad
H_{[3]}\rightarrow H_{[3]}+\Theta \s \left( G_{[4]} \wedge G_{[4]} \right)\,,
\end{equation}
where $F_{[2]}$ and $G_{[4]}$ are the corresponding field strengths of $A_{[1]}$ and $H_{[3]}$. For those deformed supergravities one could expect the existence of gauge vacua configuration satisfying,
\begin{equation}\
A_{[1]}+\Theta \s \left( F_{[2]} \wedge F_{[2]} \right)=0 \;,  \qquad
H_{[3]}+\Theta \s \left( G_{[4]} \wedge G_{[4]} \right)=0\,.
\end{equation}

In all these examples the internal gauge symmetries associated to the fields $A_{[p]}$ and/or $H_{[3]}$ of the original theories will not be destroyed after deformation, and when valued in the kernel of the deformation map the action of these fields will vanish. Since the deformation procedure here defined acts only on the specified gauge fields, there might be symmetries of the undeformed theories associated to the remaining fields which might be broken after the deformation procedure takes place.
 In spite of this, the deformed theories are worth studying since they will contain non-trivial but locally undetectable vacua configurations of the gauge fields.

\section{Summary of results and conclusions}

We have introduced a general method for deforming gauge theories in three dimensions which provides models of gauge systems with a degenerated  vacua being either trivial or self-dual. We have also suggested some generalizations of this method to higher dimensional theories.

 In three dimensions the deformation consists of a non-trivial gauge field redefinition,
\be \label{defconc}
A\rightarrow A^\Theta:= A+\Theta \s F , \qquad F:=dA+A\w A \, ,
\ee
yielding new theories respecting diffeomorphism and gauge invariance. As a result, new vacua appear, besides the trivial  $A=0$,  which are those configurations for which  $A=-\Theta \s F$, corresponding to the self-dual fields introduced in \cite{Townsend:1983xs}.  
 The most remarkable features of these models is that self-dual fields are massive and at the same time  locally undetectable by charged fields and the background geometry. This means that the coupling to matter and the Rosenfeld  energy momentum tensor vanish for self-dual fields. Hence they do not back react on the spacetime geometry and charged matter fields behave as free up to geometric phases. 
 
The topological role of self-dual fields in $2+1D$ Minkowski spacetime was confirmed invoking  self-duality of  Yang-Mills fields in a  $2+2D$-signature spacetime and showing that the $2+1D$ self-duality  equations of motion are consistent as mixed boundary conditions. We have shown that this system possesses Abrikosov-like vortices, with London penetration depth provided by the deformation parameter $\Theta$ and an exponential damping in the axial direction with the same depth of penetration.  Hence we argue that these models  describe an abstract superconductor in $2+2D$.

We also hinted at the generalization of the deformation of gauge theories to higher dimensions and to higher order gauge forms. We wrote an expression  that generalizes \eqref{defconc},
\be\label{sdgeneralc}
A_{[p]} \quad \rightarrow  \quad A^{\bf \Theta} :=A_{[p]}+ f_{[p]}({\bf \Theta}; F_{[p+1]}\, ; \s )=0 \,,
\ee
were the shift $f_{[p]}({\bf \Theta}; F_{[p+1]}\, ; \s ) $ is a $p$-form constructed with the gauge curvature tensor $F_{[p+1]}$, the spacetime metric and a set of constant parameters ${\bf \Theta}$. The deformed theory will possess vacua configurations,
\be\label{sdgeneralc}
A_{[p]}+ f_{[p]}({\bf \Theta}; F_{[p+1]}\, ; \s )=0 \,.
\ee
We  argued that they may exhibit similar features as the deformed theories in three dimensions. Those systems may be of interest since they may realize states with non-trivial degrees of freedom condensing to locally undetectable configurations.
It would be interesting to look for the analogous of the Abrikosov-like  vortex configurations found in three dimensions for which we expect satisfy equation \eqref{sdgeneralc}. In particular this strategy can be applied for the one-form and three-form  $A_{[1]}$ and $H_{[3]}$ of $D=5$ and $D=11$ supergravity, respectively.

 The deformation of Chern-Simons theory coupled to matter yields at first order of approximation in the deformation parameter a theory which was studied in detail in reference \cite{Itzhaki:2002rc}. There it was shown that the generalized Wilson loop
\be \label{Wloop}
\mathcal{P}( \exp \oint A^\Theta )\, ,
\ee
creates a charge-flux bound state, known as an anyon, which exhibits fractional statistics. This result suggests that the deformation of the theories in $2+1D$ may induce transmutation of statistics with respect to the undeformed system.

It is worth mentioning also  that generally non-minimally coupled scalars fields with vanishing energy momentum tensor were explored in \cite{AyonBeato:2004ig} and \cite{AyonBeato:2005tu} where they were dubbed as \textit{stealth fields}. It was actually shown in \cite{Hassaine:2006gz} that, up to boundary terms, the action for this system has a lower and upper bound when evaluated on BTZ black hole, which are both saturated and coincide, for the stealth configuration, yielding the vanishing of the action.\footnote{For the particular case of a conformally coupled scalar in four dimensions the possible relation between self-duality and stealth behavior was explored in \cite{de Alfaro:1976qz}.} Hence the scalar stealth systems exhibit similar features than the self-dual fields introduced here.

\paragraph*{Acknowledgments }

We thank Eloy Ayon-Beato, Patricia Ritter and Jorge Zanelli for discussions. This work has been supported by FONDECYT grants N$^{\rm o}$ 3120103 and N$^{\rm o}$ 1141073.

\end{document}